\documentclass[preprint,showpacs]{revtex4}%
\usepackage[T2A]{fontenc}
\usepackage[cp1251]{inputenc}
\usepackage[english]{babel}
\usepackage{amsmath}
\usepackage{amsfonts}
\usepackage{graphicx,epsfig}
\usepackage{subfigure}
\usepackage{amssymb}
\usepackage[indent,small]{caption2}
\usepackage{graphicx}%
\providecommand{\U}[1]{\protect\rule{.1in}{.1in}}
\begin{document}

\title{Optical eigenmodes in plane arrays of cylindrical waveguides. Analysis by means of
multiple Mie scattering formalism and phenomenological model.}

\author{Gozman M. I.$^1$, Polishchuk Yu. I.$^2$, Habarova T. V.$^1$, Polishchuk I. Ya.$^{1,2}$}

\affiliation{$^1$ RRC Kurchatov Institute, Kurchatov Sq., 1,
123182 Moscow, Russia}

\affiliation{$^2$ Moscow Institute of Physics and Technology,
141700, 9, Institutskii per., Dolgoprudny, Moscow Region, Russia}

\begin{abstract}

We consider a plane periodical array of parallel cylindrical
waveguides with evanescent coupling between them. A new method for
calculating of isofrequency curves based on the multiple Mie
scattering formalism (MMSF) is developed. This method is compared
with the phenomenological model. The derivation of the
phenomenological model by means of the MMSF is performed. The
formulae for calculation of parameters of the phenomenological
model are derived, such as propagation constants and coupling
constant.

\end{abstract}

\maketitle


\section{Introduction.} \label{Sec:Intro}

Nowadays, much attention is devoted to periodical arrays of
evanescently coupled optical waveguides. Such systems represent
the particular case of low-dimensional photonic crystal
structures. The general feature of such systems is the existence
of photonic band structure \cite{01_Joannopoulos1997,02_Busch2004}
that is analogous to the electron band structure in solids.
Therefore some effects in photonic crystal structures may be
analogous to some phenomena in solids \cite{03_Longhi2009}.

In this paper we consider a plane array of parallel equidistant
waveguides. We assume that the interaction between the waveguides
is enough weak but not negligible. In this case, the eigenmodes of
the array may be represented in a spirit of tight binding method
taken from the solid state physics. It means that the eigenmodes
of the array can be expressed in terms of the eigenmodes of the
noninteracting waveguides.


The eigenmodes for the $j$-th waveguide are described as follows
\cite{04_Marcuse1972}

\begin{equation}
\begin{array}[c]{c} \displaystyle
\mathbf{E}_j(\mathbf{r})=e^{-i\omega t+i\beta_j
z}\,\mathbf{U}_j(x_j,y_j),
\medskip \\ \displaystyle
\mathbf{H}_j(\mathbf{r})=e^{-i\omega t+i\beta_j
z}\,\mathbf{V}_j(x_j,y_j),
\end{array}
\label{EigModeSingleWG}
\end{equation}

where $\omega$ is a frequency of an eigenmode, $x_j$, $y_j$, $z$
are the coordinates of a point $\mathbf{r}$ with respect to the
axis of the waveguide, $\beta_j$ is the propagation constant of
the $j$-th waveguide. If $\beta_j>\omega$ (the speed of light is
assumed to be unit), the functions $\mathbf{U}_j(x_j,y_j)$,
$\mathbf{V}_j(x_j,y_j)$ outside the waveguide decrease
exponentially as the distance of the observation point from the
waveguide increases. Thus, the mode is evanescent and it cannot be
converted into a free photon.

So, the eigenmodes of the array of weakly interacting waveguides
may be represented in following way:

\begin{equation}
\begin{array}[c]{c} \displaystyle
\mathbf{E}(\mathbf{r})\approx e^{-i\omega t}\sum\limits_{j=1}^N
A_j(z)\,\mathbf{U}_j(x_j,y_j),
\medskip \\ \displaystyle
\mathbf{H}(\mathbf{r})\approx e^{-i\omega t}\sum\limits_{j=1}^N
A_j(z)\,\mathbf{V}_j(x_j,y_j).
\end{array}
\label{CollectiveMode}
\end{equation}

If the distance between the waveguides is large enough, the
coupling between only the nearest waveguides may be taken into
account. Then, the equation for an eigenmode of the array reads

\begin{equation}
i\frac{dA_j}{dz}(z)+\beta_j A_j(z)+
\gamma\Bigl(A_{j-1}(z)+A_{j+1}(z)\Bigr)=0, \label{Tradi}
\end{equation}

where $\gamma$ is a nearest neighbor coupling constant (for
derivation see, for example, \cite{04_Marcuse1972}). This equation
or the analogous equations are usually used for simulation of
optical effects in systems of interacting waveguides, such as
optical Bloch oscillations
\cite{05_Pertsch1998,06_Pertsch1999,07_Pertsch2002}, Zener
tunneling \cite{08_Pertsch2006,09_Pertsch2009,10_ZhengWang2010},
dynamic localization \cite{11_Longhi2005,12_Longi2006}, Anderson
localization \cite{15_Lahini2008}.

A principle drawback of Eq. (\ref{Tradi}) is that the
phenomenological constants $\beta_j$ and $\gamma$ are unknown.
They can be found from the experiment if one supposes that Eq.
(\ref{Tradi}) is applicable to the system under study.

However, in the important particular case of the cylindrical
waveguides, one may propose the exact description of the optical
properties of the array. In this case every eigenmode is
characterized by the angular momentum $m$. The eigenmodes are
described as follows \cite{04_Marcuse1972}

\begin{equation}
\begin{array}[c]{c} \displaystyle
\mathbf{E}_{jm}(\mathbf{r})=e^{-i\omega
t+im\phi_j+i\beta_{jm}z}\,\mathbf{U}_{jm}(\rho_j),
\medskip \\ \displaystyle
\mathbf{H}_{jm}(\mathbf{r})=e^{-i\omega
t+im\phi_j+i\beta_{jm}z}\,\mathbf{V}_{jm}(\rho_j).
\end{array}
\label{EigModeSingleCylWG}
\end{equation}

Here $m$ is the angular momentum, $\rho_j$, $\phi_j$ are polar
coordinates of a point $\mathbf{r}$ with respect to the axis of
the waveguide, $U_{jm}(\rho_j),~V_{jm}(\rho_j)\sim
H_m(\varkappa_{jm}\rho_j)$, where $H_m(x)$ is a Hankel function of
the first kind, $\varkappa_{jm}=\sqrt{\omega^2-\beta_{jm}^2}$.

The rigorous formalism for description of array of cylindrical
waveguides makes use of the exact solution for the electromagnetic
wave scattering problem by an infinite cylinder. The proposed
description is based on the possibility to generalize this
solution for the case of many parallel cylinders --- multiple Mie
scattering formalism (MMSF) for the arrays of infinite cylinders
\cite{16_VanDeHulst1981,17_Felbacq1994,18_ZhangLi1998,19_Felbacq2000,20_DuLin2009}.
This approach is similar to the multiple Mie scattering formalism
for spherical particles
\cite{21_YuLinXu1998,22_We2008,23_We2009,24_We2010}. The MMSF can
be used for investigation of scattering and transmission of light
by photonic crystals \cite{18_ZhangLi1998,19_Felbacq2000} or
irregular systems of cylindrical waveguides \cite{17_Felbacq1994},
for calculation of the eigenmode frequencies and band structures
of a plane array of cylindrical waveguides \cite{20_DuLin2009} and
of two-dimensional photonic crystals \cite{19_Felbacq2000}.

In this paper we use the MMSF for calculation of the isofrequency
curves for the array of identical waveguides. We consider the case
when the waveguides of the array are situated close to each other
and ascertain if the phenomenological approach is applicable for
such a system. For this purpose we derive the phenomenological
approach from the MMSF and develop the method for calculation the
coupling constant.

\bigskip

The MMSF is explained in Sect. \ref{Sec:MMSF}. In Sect.
\ref{Sec:Relationship} we discuss the connection between the MMSF
and the phenomenological approach and explain how the coupling
constant $\gamma$ in Eq. (\ref{Tradi}) can be calculated. In Sect.
\ref{Sec:Application} we calculate the isofrequency curves of a
plane array of infinite cylindrical waveguides. We compare the
isofrequency curves calculated by means of MMSF with that
calculated by the phenomenological model. In Conclusion we discuss
the possibility of further development of method used in this
paper.


\section{Multiple Mie scattering formalism.}
\label{Sec:MMSF}

Let us consider an array of $N$ parallel cylindrical dielectric
rods. The axes of the rods are in the plane $y=0$ and they all are
parallel to the $z$-axis. The refractive index of the $j$-th array
is denoted as $n_j$. Let the array is illuminated by the external
field of the certain frequency $\omega$ and longitudinal wave
vector $\beta$:

\begin{equation}
\mathbf{E}^\text{ext}(\mathbf{r}) =e^{-i\omega t+i\beta
z}~\mathbf{E}^\text{ext}(x,y), \qquad
\mathbf{H}^\text{ext}(\mathbf{r})=e^{-i\omega t+i\beta
z}~\mathbf{H}^\text{ext}(x,y). \label{MSF010}
\end{equation}

This field causes the response of the array. The field inside the
$j$-th rod is

\begin{equation}
\begin{array}[c]{l} \displaystyle
\tilde{\mathbf{E}}_j(\mathbf{r})=e^{-i\omega t+i\beta z}
\sum\limits_m e^{im\phi_j}\Bigl(c_{jm}\,\mathbf{M}_{\omega_j \beta
m}^1 (\rho_j)-d_{jm}\,\mathbf{N}_{\omega_j\beta
m}^{1}(\rho_j)\Bigr),
\medskip \\ \displaystyle
\tilde{\mathbf{H}}_j(\mathbf{r})=e^{-i\omega t+i\beta z}
\sum\limits_m e^{im\phi_j}\Bigl(c_{jm}\,\mathbf{N}_{\omega_j\beta
m}^1 (\rho_j)+d_{jm}\,\mathbf{M}_{\omega_j\beta
m}^1(\rho_j)\Bigr), \qquad \rho_j<R.
\end{array}
\label{AppS_EHint}
\end{equation}

Here $\rho_j$, $\phi_j$, $z$ are the cylindrical coordinates of
the point $\mathbf{r}$ respectively to the axis of the $j$-th
waveguide, and $\omega_j=n_j\omega$. The functions
$\mathbf{M}_{\omega_j\beta m}^1(\rho_j)$ and
$\mathbf{N}_{\omega_j\beta m}^1(\rho_j)$ are linear superpositions
of the Bessel functions. The partial amplitudes $c_{jm}$, $d_{jm}$
determine the field inside the $j$-th rod. Below the factor
$e^{-i\omega t+i\beta z}$ is omitted, for short.

The field scattered by the $j$-th rod may be represented in the
form

\begin{equation}
\begin{array}[c]{l} \displaystyle
\mathbf{E}_j(\mathbf{r})=\sum\limits_m e^{im\phi_j}~\Bigl(a_{jm}~
\mathbf{M}_{\omega\beta
m}^2(\rho_j)-b_{jm}~\mathbf{N}_{\omega\beta m}^2(\rho_j)\Bigr),
\medskip \\ \displaystyle
\mathbf{H}_j(\mathbf{r})=\sum\limits_m e^{im\phi_j}~\Bigl(a_{jm}
~\mathbf{N}_{\omega\beta
m}^2(\rho_j)+b_{jm}~\mathbf{M}_{\omega\beta m}^2(\rho_j)\Bigr),
\qquad \rho_j>R.
\end{array}
\label{EX_J}
\end{equation}

The functions $\mathbf{M}_{\omega\beta m}^2(\rho_j)$ and
$\mathbf{N}_{\omega\beta m}^2(\rho_j)$ are linear superpositions
of the Hankel functions of the first kind for the imaginary
argument.

On the other hand, the field Eq. (\ref{EX_J}) can be represented
in the alternative form as an expansion in terms of functions
$\mathbf{M}_{\omega\beta m}^1(\rho_l)$ and
$\mathbf{N}_{\omega\beta m}^1(\rho_l)$ for any $l\neq j$:

\begin{equation}
\begin{array}[c]{l} \displaystyle
\mathbf{E}_j(\mathbf{r})=\sum_m e^{im\phi_l}~\Bigl(p_{jm}^l
~\mathbf{M}_{\omega\beta m}^1(\rho_l)-
q_{jm}^l~\mathbf{N}_{\omega\beta m}^1(\rho_l)\Bigr),
\medskip \\ \displaystyle
\mathbf{H}_j(\mathbf{r})=\sum_m e^{im\phi_l}~\Bigl(p_{jm}^l~
\mathbf{N}_{\omega\beta m}^1(\rho_l)+q_{jm}^l~
\mathbf{M}_{\omega\beta m}^1(\rho_l)\Bigr), \qquad ~l\neq j.
\end{array}
\label{EX_J1}
\end{equation}

Let us emphasize that the Eqs. (\ref{EX_J}) and (\ref{EX_J1})
represent the same field, i. e. the field scattered by the $j$-th
waveguide.

According to \cite{17_Felbacq1994}, one can relate the amplitudes
$p_{jm}^l$, $q_{jm}^l$ and $a_{lm}$, $b_{lm}$ as follows

\begin{equation}
p_{jm}^l=\sum\limits_{n=-\infty}^{+\infty}\,U_{jm}^{ln}(\omega,\beta)~
a_{ln}, \qquad q_{jm}^l=\sum\limits_{n=-\infty}^{+\infty}\,
U_{jm}^{ln}(\omega,\beta)~b_{ln}, \label{pq=Uab}
\end{equation}

where

\begin{equation}
U_{jm}^{ln}(\omega,\beta) =H_{n-m}(\varkappa a\cdot\left\vert
j-l\right\vert)\times\left\{
\begin{array}[c]{cc}
1 & \text{if } l>j, \\
(-1)^{m-n} & \text{if } l<j,
\end{array}
\right.\label{U}
\end{equation}

$\varkappa=\sqrt{\omega^2-\beta^2}$ and $H_m(x)$ is the Hankel
function of the first kind.

Let us introduce a notation

\begin{equation}
\begin{array}[c]{l} \displaystyle
\mathbf{E}'_j(\mathbf{r})=\sum\limits_{l\neq
j}\mathbf{E}_l(\mathbf{r})=\sum\limits_{l\neq j}\sum\limits_m
e^{im\phi_j}~\Bigl(p_{jm}^l\mathbf{M}_{\omega\beta
m}^1(\rho_j)-q_{jm}^l\,\mathbf{N}_{\omega\beta m}^1(\rho_j)\Bigr),
\medskip \\ \displaystyle
\mathbf{H}'_j(\mathbf{r})=\sum_{l\neq
j}\mathbf{H}_l(\mathbf{r})=\sum\limits_{l\neq j}\sum\limits_m
e^{im\phi_j}~\Bigl(p_{jm}^l~\mathbf{N}_{\omega\beta
m}^1(\rho_j)+q_{jm}^l\mathbf{M}_{\omega\beta
m}^1(\rho_j)\Bigr),\qquad \rho_j>R.
\end{array}
\label{EHprime}
\end{equation}

One can rewrite it in the form

\begin{equation}
\begin{array}[c]{l} \displaystyle
\mathbf{E}'_j(\mathbf{r})=\sum_{l\neq
j}\mathbf{E}_l(\mathbf{r})=\sum\limits_m
e^{im\phi_j}~\Bigl(p_{jm}\,\mathbf{M}_{\omega\beta
m}^1(\rho_j)-q_{jm}\,\mathbf{N}_{\omega\beta m}^1(\rho_j)\Bigr),
\medskip \\ \displaystyle
\mathbf{H}'_j(\mathbf{r})=\sum_{l\neq
j}\mathbf{H}_l(\mathbf{r})=\sum\limits_m
e^{im\phi_j}~\Bigl(p_{jm}\,\mathbf{N}_{\omega\beta
m}^1(\rho_j)+q_{jm}\,\mathbf{M}_{\omega\beta
m}^1(\rho_{j})\Bigr),\qquad \rho_j>R.
\end{array}
\label{EHprime1}
\end{equation}

Here

\begin{equation}
\begin{array}[c]{l}\displaystyle
p_{jm}=\sum\limits_{l\neq j}\,p_{jm}^l=\sum_{l\neq
j}\sum\limits_{n=-\infty}^{+\infty}\,U_{jm}^{ln}(\omega,\beta)\,a_{ln},
\medskip \\ \displaystyle
q_{jm}=\sum_{l\neq j}\,q_{jm}^l=\sum_{l\neq
j}\sum\limits_{n=-\infty
}^{+\infty}\,U_{jm}^{ln}(\omega,\beta)\,b_{ln}.
\end{array}
\label{tr12}
\end{equation}

Let us assume that the external field

\begin{equation}
\begin{array}[c]{l} \displaystyle
\mathbf{E}^\text{ext}(\mathbf{r})=\sum\limits_m ~e^{im\phi_j}
\Bigl(P_m^j\,\mathbf{M}_{\omega\beta
m}^1(\rho_j)-Q_m^j\,\mathbf{N}_{\omega\beta m}^1(\rho_j)\Bigr),
\medskip \\ \displaystyle
\mathbf{H}^\text{ext}(\mathbf{r})=\sum\limits_m ~e^{im\phi_j}
\Bigl(P_m^j\,\mathbf{N}_{\omega\beta
m}^1(\rho_j)+Q_m^j\,\mathbf{M}_{\omega\beta m}^1(\rho_j)\Bigr).
\end{array}
\label{out}
\end{equation}

Then, the field outside of the array may be represented in the
form

\begin{equation}
\begin{array}[c]{l} \displaystyle
\mathbf{E}(\mathbf{r})=\mathbf{E}^\text{ext}(\mathbf{r})+
\mathbf{E}_j(\mathbf{r})+\sum\limits_{l\neq
j}\mathbf{E}_l(\mathbf{r}),
\medskip \\ \displaystyle
\mathbf{H}(\mathbf{r})=\mathbf{H}^\text{ext}(\mathbf{r})+\mathbf{H}_j(\mathbf{r})
+\sum\limits_{l\neq j}\mathbf{H}_l(\mathbf{r}),
\end{array}
\label{out1}
\end{equation}

where the number $j$ is arbitrary.

The relations between fields outside and inside the $j$-th rod
follow from the boundary conditions on its surface. These
relations take the following form:

\begin{equation}
\left(\begin{matrix} a_{mj} \\ b_{mj}
\end{matrix}\right)=\hat{S}_{jm}(\omega,\beta)~
\left(\begin{matrix}
P_m^j+p_{jm} \\
Q_m^j+q_{jm}
\end{matrix}\right),
\label{AppS_ab=Spq}
\end{equation}

\begin{equation}
\left(\begin{matrix} c_{mj} \\ d_{mj}
\end{matrix}\right)=\hat{T}_{mj}(\omega,\beta)~
\left(\begin{matrix} a_{mj} \\ b_{mj}
\end{matrix} \right).
\label{AppS_cd=Tab}
\end{equation}

Taking into account Eq. (\ref{tr12}) in Eq. (\ref{AppS_ab=Spq})
one obtains the self-consistent system of equations

\begin{equation}
\hat{S}_{jm}^{-1}(\omega,\beta)\, \left(
\begin{matrix}
a_{jm} \\ b_{jm}
\end{matrix} \right)
-\sum\limits_{l\neq
j}^N\,\sum\limits_{n=-\infty}^{+\infty}\,U_{jm}^{ln}(\omega,\beta)\,
\left(
\begin{matrix} a_{ln} \\ b_{ln}
\end{matrix} \right)
=\left( \begin{matrix} P_m^j \\ Q_m^j
\end{matrix} \right).
\label{MainSyst}
\end{equation}

The system of equation Eq. (\ref{MainSyst}) describes the response
of the array on the external electromagnetic field, determined by
the amplitudes $P_m^j$, $Q_m^j$. At the same time, if one takes
$P_m^j=Q_m^j=0$, the Eq. (\ref{MainSyst}) describes the
electromagnetic eigenmodes for the array under consideration.
These modes are described as follows:

\begin{equation}
\hat{S}_{jm}^{-1}(\omega,\beta)~\left(
\begin{matrix}
a_{jm} \\ b_{jm}
\end{matrix}\right)
-\sum\limits_{l\neq j}^N\,\sum\limits_{n=-\infty}^{+\infty}
\,U_{jm}^{ln}(\omega,\beta)\, \left(
\begin{matrix} a_{ln} \\ b_{ln}
\end{matrix} \right)  =0.
\label{MainSyst*}
\end{equation}

The homogeneous system (\ref{MainSyst*}) possesses a nontrivial
solution only if

\begin{equation}
\det~\left\vert
\hat{S}_{jm}^{-1}(\omega,\beta)~\delta_{jl}~\delta_{mn}-U_{jm}^{ln}(\omega,\beta)\right\vert=0.
\label{det=0}
\end{equation}

This equation allows to obtain the eigenvalues of $\beta$ for the
eigenmodes of the array.

In particular, for the single rod this equation takes the form

\begin{equation}
\det~\left\vert\hat{S}_{jm}^{-1}(\omega,\beta)\right\vert=0.
\label{det1=0}
\end{equation}

The solutions of this equation $\beta_{jm}$ (depending on
$\omega$) are the propagation constants of the $j$-th waveguide,
that is assumed noninteracting with the other waveguides. One can
see that these propagation constants are characterized by the
angular momentum $m$, as it was mentioned above.

Below we apply Eq. (\ref{MainSyst}) and Eq. (\ref{MainSyst*}) to
describe the optical properties of the array of the rods.


\section{Relationship of the multiple scattering formalism and the phenomenological approach.}
\label{Sec:Relationship}

Let us derive the simplified equations which describe the optical
properties of the array of the rods under consideration.

Every rod is characterized by a set of its propagation constants
$\beta_{jm}$, satisfying to Eq. (\ref{det1=0}). Let us notice that
the propagation constants corresponding to the opposite angular
momenta coincide, i. e. $\beta_{jm}=\beta_{j,-m}$.

We suppose that the propagation constants of different waveguides
differ slightly. Besides, the coupling $U_{jm}^{ln}(\omega,\beta)$
is weak and may be considered as a perturbation with respect to
$\hat{S}_{jm}^{-1}(\omega,\beta)$. Therefore, we can consider the
optical excitations originated from the propagation constants with
fixed angular momentum $m$. Two cases are possible: $m=0$ and
$m\neq 0$. For the first case one should take into account two
partial amplitudes $a_{j0}$, $b_{j0}$. For the second case the
system of equations should include four partial amplitudes
$a_{jm}$, $b_{jm}$ and $a_{j,-m}$, $b_{j,-m}$, since the
propagation constants for the angular momenta $m$ and $-m$
coincide.

Below we take into account only the coupling between the nearest
neighbors, since the coupling is evanescent.


\subsection{First case: \qquad  $m=0$.}

The first case is $m=0$. In this case the main system of equations
takes the form

\begin{equation}
\hat{S}_{j0}^{-1}(\omega,\beta)\,\left(\begin{matrix} a_{j0}
\\ b_{j0}
\end{matrix}\right)
-\sum\limits_{l=j\pm1}~U_{j0}^{l0}(\omega,\beta)\,
\left(\begin{matrix} a_{l0} \\ b_{l0}
\end{matrix}\right)=0. \label{G010}
\end{equation}

Below we omit the arguments $\omega$, $\beta$ for short.

The matrix $\hat{S}_{j0}^{-1}$ is diagonal,

\begin{equation}
\hat{S}_{j0}^{-1}=\left(\begin{matrix} A_j & 0 \\ 0 & B_j
\end{matrix}\right). \label{G020}
\end{equation}

Here $A_j$, $B_j$ are some functions of $\omega$ and $\beta$. So,
Eq. (\ref{G010}) separates in two independent systems of
equations:

\begin{equation}
A_j~a_{j0}-\sum\limits_{l=j\pm1}~U_{j0}^{l0}~a_{l0}=0,
\label{G030a}
\end{equation}

\begin{equation}
B_j~b_{j0}-\sum\limits_{l=j\pm1}~U_{j0}^{l0}~b_{l0}=0,
\label{G030b}
\end{equation}

Each propagation constant $\beta_{j0}$ satisfy to one of the
following equations:

\begin{equation}
A_j(\omega,\beta_{j0})=0, \qquad B_j(\omega,\beta_{j0})=0.
\label{G040}
\end{equation}

Below we consider Eq. (\ref{G030a}) only, since for Eq.
(\ref{G030b}) the derivation is similar.

Since the coupling between the waveguides is weak, the
isofrequency curve originating from any propagation constant is
narrow. Therefore one can represent $A_j$ in following way:

\begin{equation}
A_j=D^A_j\times(\beta-\beta_{j0}). \label{G050}
\end{equation}

Here $\beta_{j0}$ satisfies to the first of Eqs. (\ref{G040}).

Substituting this to Eq. (\ref{G030a}), we obtain:

\begin{equation}
(\beta-\beta_{j0})~a_{j0}-\sum\limits_{l=j\pm1}~\frac{U_{j0}^{l0}}{D^A_j}
~a_{l0}=0. \label{G060}
\end{equation}

Let us notice that $U_{j0}^{j-1,0}=U_{j0}^{j+1,0}$. Here we assume
that $U_{j0}^{l0}(\omega,\beta)=U_{j0}^{l0}(\omega,\beta_{j0})$
and that the value $U_{j0}^{j\pm1,0}(\omega,\beta_{j0})/D^A_j$ is
independent on $j$. So, introducing the notation

\begin{equation}
\gamma=\frac{U_{j0}^{j\pm1,0}(\omega,\beta_{j0})}{D^A_j},
\label{G065}
\end{equation}

we obtain

\begin{equation}
(\beta-\beta_{j0})~a_{j0}-\gamma\,\Bigl(a_{j-1,0}+a_{j+1,0}\Bigr)=0.
\label{G070}
\end{equation}

The Eq. (\ref{G070}) possesses the nontrivial solutions only for
eigenvalues of $\beta$.

The electric field outside the array of waveguides is

\begin{equation}
\mathbf{E}(t,\mathbf{r})=e^{-i\omega t}~\sum\limits_{j=1}^N
\sum\limits_\beta ~e^{i\beta z}~ a_{j0}(\beta)~
\mathbf{M}_{\omega\beta 0}^2(\rho_j). \label{G071}
\end{equation}

The expression for the magnetic field is analogous.

Here $\sum\limits_\beta$ means the sum over the eigenvalues of
$\beta$. We have added the argument $\beta$ to the partial
amplitudes $a_{j0}$, $b_{j0}$, since the partial amplitudes depend
on the eigenvalue $\beta$.

Since the rods differ slightly and the interaction between them is
weak, all the eigenvalues of $\beta$ are close to each other. So,
one can suppose that $\mathbf{M}_{\omega\beta
0}^2(\rho_j)=\mathbf{M}_{\omega\beta_{j0} 0}^2(\rho_j)$. Let us
introduce the notation

\begin{equation}
a_{j0}(z)=\sum\limits_\beta ~e^{i\beta z}~ a_{j0}(\beta).
\label{G072}
\end{equation}

So, the equation (\ref{G071}) takes the form:

\begin{equation}
\mathbf{E}(t,\mathbf{r})=e^{-i\omega t}~\sum\limits_{j=1}^N
~a_{j0}(z)~ \mathbf{M}_{\omega\beta_{j0} 0}^2(\rho_j).
\label{G073}
\end{equation}

Taking into account Eq. (\ref{G070}), one can write the equation
for $a_{j0}(z)$:

\begin{equation}
\left(i\frac{d}{dz}+\beta_{j0}\right)~a_{j0}(z)+
\gamma~\Bigl(a_{j-1,0}(z)+a_{j+1,0}(z)\Bigr)=0. \label{G073}
\end{equation}

This equation coincides to Eq. (\ref{Tradi}).

In the similar way one can derive the equation

\begin{equation}
\left(i\frac{d}{dz}+\beta_{j0}\right)~b_{j0}(z)+
\gamma~\Bigl(b_{j-1,0}(z)+b_{j+1,0}(z)\Bigr)=0. \label{G073}
\end{equation}

where

\begin{equation}
\gamma=\frac{U_{j0}^{j\pm1,0}(\omega,\beta_{j0})}{D^B_j},
\label{G074}
\end{equation}

and $D^B_j$ is determined by the equation

\begin{equation}
B_j=D^B_j\times(\beta-\beta_{j0}). \label{G075}
\end{equation}


\subsection{Second case: \qquad $m\neq 0$.}

The second case is $m\neq 0$. As it was mentioned above, we should
take into account the partial amplitudes $a_{jm}$, $b_{jm}$ and
$a_{j,-m}$, $b_{j,-m}$ both. So, the main system of equations
takes the form

\begin{equation}
\begin{array}{c} \displaystyle
\hat{S}_{jm}^{-1}~\left(\begin{matrix} a_{jm}
\\ b_{jm} \end{matrix}\right)-\sum\limits_{l=j\pm
1}~\left\{U_{jm}^{lm}~\left(\begin{matrix} a_{lm}
\\ b_{lm} \end{matrix}\right)+U_{jm}^{l,-m}~\left(\begin{matrix} a_{l,-m}
\\ b_{l,-m} \end{matrix}\right)\right\}=0,
\bigskip \\ \displaystyle
\hat{S}_{j,-m}^{-1}~\left(\begin{matrix} a_{j,-m}
\\ b_{j,-m} \end{matrix}\right)-\sum\limits_{l=j\pm
1}~\left\{U_{j,-m}^{l,-m}~\left(\begin{matrix} a_{l,-m} \\
b_{l,-m}
\end{matrix}\right)+U_{j,-m}^{lm}~\left(\begin{matrix}
a_{lm} \\ b_{lm} \end{matrix}\right)\right\}=0.
\end{array}
\label{G080}
\end{equation}

It is convenient to introduce the following notations:

\begin{equation}
\begin{array}{c} \displaystyle
U_{jm}^{j-1,m}=U_{jm}^{j+1,m}=U_{j,-m}^{j-1,-m}=U_{j,-m}^{j+1,-m}=U_m,
\medskip \\ \displaystyle
U_{jm}^{j-1,-m}=U_{jm}^{j+1,-m}=U_{j,-m}^{j-1,m}=U_{j,-m}^{j+1,m}=V_m.
\end{array}
\label{G090}
\end{equation}

Substituting this to Eq. (\ref{G080}), one gets

\begin{equation}
\begin{array}{c} \displaystyle
\hat{S}_{jm}^{-1}~\left(\begin{matrix} a_{jm}
\\ b_{jm} \end{matrix}\right)-\sum\limits_{l=j\pm
1}~\left\{U_m~\left(\begin{matrix} a_{lm}
\\ b_{lm} \end{matrix}\right)+V_m~\left(\begin{matrix} a_{l,-m}
\\ b_{l,-m} \end{matrix}\right)\right\}=0,
\bigskip \\ \displaystyle
\hat{S}_{j,-m}^{-1}~\left(\begin{matrix} a_{j,-m}
\\ b_{j,-m} \end{matrix}\right)-\sum\limits_{l=j\pm
1}~\left\{U_m~\left(\begin{matrix} a_{l,-m}
\\ b_{l,-m} \end{matrix}\right)+V_m~\left(\begin{matrix} a_{lm}
\\ b_{lm} \end{matrix}\right)\right\}=0.
\end{array}
\label{G100}
\end{equation}

Below we show that there are two types of solutions.

Let us suppose that the partial amplitudes $a_{jm}$, $b_{jm}$ and
$a_{j,-m}$, $b_{j,-m}$ are connected by the following relation:

\begin{equation}
\left(\begin{matrix} a_{j,-m}
\\ b_{j,-m} \end{matrix}\right)=\hat{M}~\left(\begin{matrix} a_{jm}
\\ b_{jm} \end{matrix}\right).
\label{G110}
\end{equation}

Substituting this to (\ref{G100}), one gets:

\begin{equation}
\begin{array}{c} \displaystyle
\hat{S}_{jm}^{-1}~\left(\begin{matrix} a_{jm}
\\ b_{jm} \end{matrix}\right)-\sum\limits_{l=j\pm
1}~\Bigl(U_m+V_m\,\hat{M}\Bigr)~\left(\begin{matrix} a_{lm} \\
b_{lm} \end{matrix}\right)=0,
\bigskip \\ \displaystyle
\hat{M}^{-1}~\hat{S}_{j,-m}^{-1}\hat{M}~\left(\begin{matrix}
a_{jm} \\ b_{jm} \end{matrix}\right)-\sum\limits_{l=j\pm
1}~\Bigl(U_m+V_m\,\hat{M}^{-1}\Bigr)~\left(\begin{matrix} a_{lm}
\\ b_{lm} \end{matrix}\right)=0.
\end{array}
\label{G110}
\end{equation}

Both equations in (\ref{G110}) should coincide. Therefore the
matrix $\hat{M}$ should satisfy to following conditions:

\begin{equation}
\begin{array}{c} \displaystyle
\hat{M}^{-1}=\hat{M},
\bigskip \\ \displaystyle
\hat{M}^{-1}~\hat{S}^{-1}_{j,-m}~\hat{M}=\hat{S}^{-1}_{jm}.
\end{array}
\label{G120}
\end{equation}

To find the possible forms of matrix $\hat{M}$ one should use a
relation between the matrices $\hat{S}_{jm}$ and $\hat{S}_{j,-m}$.
These matrices possess the form

\begin{equation}
\hat{S}^{-1}_{jm}=\left(\begin{matrix} iA & ~~C \\ -C & ~~iB
\end{matrix}\right), \qquad
\hat{S}^{-1}_{j,-m}=\left(\begin{matrix} iA & ~~-C \\ C & ~~iB
\end{matrix}\right),
\label{G130}
\end{equation}

where $A$, $B$ and $C$ are some real functions of $\omega$ and
$\beta$.

So, one can find easily, that there are only two possible forms of
the matrix $\hat{M}$:

\begin{equation}
\hat{M}=\left(\begin{matrix} 1 & ~~0 \\ 0 & ~~-1
\end{matrix}\right), \qquad \text{or} \qquad \hat{M}=\left(\begin{matrix} -1 & ~~0 \\ 0 &
~~1 \end{matrix}\right). \label{G140}
\end{equation}

So, we see that the solutions of equation (\ref{G080}) separates
in two different types.

For the solutions of the first type

\begin{equation}
a_{j,-m}=a_{jm}, \qquad b_{j,-m}=-b_{jm}, \label{G149}
\end{equation}

and $a_{jm}$, $b_{jm}$ satisfy the equation

\begin{equation}
\hat{S}^{-1}_{jm}~\left(\begin{matrix} a_{jm} \\
b_{jm} \end{matrix}\right)-\sum\limits_{l=j\pm 1}~
\left(\begin{matrix} U_m+V_m & 0 \\ 0 & U_m-V_m
\end{matrix} \right)~\left(\begin{matrix} a_{lm} \\ b_{lm}
\end{matrix}\right)=0. \label{G150}
\end{equation}

For the solutions of the second type

\begin{equation}
a_{j,-m}=-a_{jm}, \qquad b_{j,-m}=b_{jm}, \label{G159}
\end{equation}

and $a_{jm}$, $b_{jm}$ satisfy the equation

\begin{equation}
\hat{S}^{-1}_{jm}~\left(\begin{matrix} a_{jm} \\
b_{jm} \end{matrix}\right)-\sum\limits_{l=j\pm 1}~
\left(\begin{matrix} U_m-V_m & 0 \\ 0 & U_m+V_m
\end{matrix} \right)~\left(\begin{matrix} a_{lm} \\ b_{lm}
\end{matrix}\right)=0. \label{G160}
\end{equation}

\bigskip

Below we consider an equation

\begin{equation}
\hat{S}^{-1}_{jm}~\left(\begin{matrix} a_{jm} \\
b_{jm} \end{matrix}\right)-\sum\limits_{l=j\pm 1}~
\hat{W}_m~\left(\begin{matrix} a_{lm} \\
b_{lm} \end{matrix}\right)=0, \label{G161}
\end{equation}

where

\begin{equation}
\begin{array}{l} \displaystyle
\hat{W}_m=\left(\begin{matrix} U_m+V_m & 0
\\ 0 & U_m-V_m \end{matrix} \right)\qquad \text{for the first
case},
\bigskip \\ \displaystyle
\hat{W}_m=\left(\begin{matrix} U_m-V_m & 0
\\ 0 & U_m+V_m \end{matrix} \right)\qquad \text{for the
second case}.
\end{array}
\label{G162}
\end{equation}

\bigskip

Let $\mathbf{u}_{jm}=\left(\begin{matrix} \tilde{a}_{jm} \\
\tilde{b}_{jm} \end{matrix}\right)$ be the solution of the
equation

\begin{equation}
\hat{S}_{jm}^{-1}(\omega,\beta_{jm})~\mathbf{u}_{jm}=0.
\label{G170}
\end{equation}

Remain that $\beta_{jm}$ satisfies to the equation
$~~\det\,\hat{S}_{jm}^{-1}(\omega,\beta_{jm})=0$. The vector
$\mathbf{u}_{jm}$ is one of the two eigenvectors for matrix
$\hat{S}_{jm}^{-1}(\omega,\beta_{jm})$ possessing a vanishing zero
eigenvalue. Let $\mathbf{v}_{jm}$ be the other eigenfunction for
the matrix $\hat{S}_{jm}^{-1}(\omega,\beta_{jm})$, with $\mu_{jm}$
being the corresponding eigenvalue. Thus,

\begin{equation}
\hat{S}_{jm}^{-1}(\omega,\beta_{jm})~\mathbf{v}_{jm}=
\mu_{jm}\,\mathbf{v}_{jm}. \label{G180}
\end{equation}

The vectors $\mathbf{u}_{jm}$, $\mathbf{v}_{jm}$ are linearly
independent. One assumes that
$\mathbf{u}_{jm}^{\dag}\,\mathbf{u}_{jm}=\mathbf{v}_{jm}^{\dag}\,\mathbf{v}_{jm}=1$.
Let us find a solution of Eq.(\ref{G150}) in the form

\begin{equation}
\left(\begin{matrix} a_{jm} \\ b_{jm}
\end{matrix}\right)=A_{jm}\,\mathbf{u}_{jm}+B_{jm}\,\mathbf{v}_{jm}.
\label{G190}
\end{equation}

Since the coupling of the adjacent waveguides is a small
perturbation to $\hat{S}_{jm}^{-1}$ in Eq. (\ref{G150}), the
vector $\left(\begin{matrix} a_{jm} \\ b_{jm}
\end{matrix}\right)$ is practically ``parallel'' to $\mathbf{u}_{jm}$. For this reason,
$|B_{jm}|\ll|A_{jm}|$. Within the perturbation approach, the value
$\beta-\beta_{jm}$ is a small parameter. Then,

\begin{equation}
\hat{S}_{jm}^{-1}(\omega,\beta)\approx\hat{S}_{jm}^{-1}(\omega,\beta_{jm})+
(\beta-\beta_{jm})\,\hat{D}_{jm}, \label{G200}
\end{equation}

where $\hat{D}_{jm}$ is the derivative of the matrix
$\hat{S}_{jm}^{-1}(\omega,\beta)$ taken at the point
$\beta=\beta_{jm}$. Substituting (\ref{G190}) and (\ref{G200}) to
(\ref{G161}), one gets:

\begin{equation}
\begin{array}{c} \displaystyle
\biggl\{\hat{S}_{jm}^{-1}(\omega,\beta_{jm})+(\beta-\beta_{jm})\,\hat
{D}_{jm}\biggr\}~\Bigl(A_{jm}\,\mathbf{u}_{jm}+B_{jm}\,
\mathbf{v}_{jm}\Bigr)-
\medskip \\ \displaystyle
-\sum\limits_{l=j\pm1}~\hat{W}_m~
\Bigl(A_{lm}\,\mathbf{u}_{lm}+B_{lm}\,\mathbf{v}_{lm}\Bigr)=0.
\end{array}
\label{G210}
\end{equation}

With Eq. (\ref{G170}) being taken into account, the first order
perturbation approach gives

\begin{equation}
\mu_{jm}\,B_{jm}\,\mathbf{v}_{jm}+(\beta-\beta_{jm})\,A_{jm}\,\hat{D}_{jm}\,
\mathbf{u}_{jm}-\sum\limits_{l=j\pm
1} A_{lm}\,\hat{W}_m\,\mathbf{u}_{jm}=0. \label{G220}
\end{equation}

It is convenient to introduce a vector $\mathbf{w}_{jm}$
completely defined by the conditions:

\begin{equation}
\begin{array}{c}
\displaystyle
\mathbf{w}_{jm}^{\dag}\,\hat{D}_{jm}\,\mathbf{u}_{jm}=1,
\medskip \\ \displaystyle
\mathbf{w}_{jm}^{\dag}\,\mathbf{v}_{jm}=0.
\end{array}
\label{G230}
\end{equation}

Multiplying Eq. (\ref{G220}) by $\mathbf{w}_{jm}^{\dag}$ results
in the equation

\begin{equation}
(\beta-\beta_{jm})\,A_{jm}-\sum\limits_{l=j\pm 1}
A_{lm}~\mathbf{w}_{jm}^{\dag}\,\hat{W}_m(\beta)\,\mathbf{u}_{jm}=0,
\label{G240}
\end{equation}

If the variation of $\beta_{jm}$ is small as $j$ changes, the
variation of the product
$\mathbf{w}_{jm}^{\dag}\,\hat{W}_m(\beta)\,\mathbf{u}_{jm}$ is
small, as well. Therefore, one can neglect its dependence on $j$.
Denoting

\begin{equation}
\gamma=\mathbf{w}_{jm}^{\dag}\,\hat{W}_m(\beta)\,\mathbf{u}_{jm}.
\end{equation}

one obtains:

\begin{equation}
(\beta-\beta_{jm})~A_{jm}-\gamma~\Bigl(A_{j-1,m}+A_{j+1,m}\Bigr)=0.
\label{G250}
\end{equation}

The electric field outside the array is

\begin{equation}
\begin{array}{r} \displaystyle
\mathbf{E}(t,\mathbf{r})=e^{-i\omega t}~\sum\limits_{j=1}^N
\sum\limits_\beta~e^{i\beta
z}~\Bigl\{e^{im\phi_j}~\Bigl(a_{jm}(\beta)
\mathbf{M}^2_{\omega\beta m}(\rho_j)-b_{jm}(\beta)
\mathbf{N}^2_{\omega\beta m}(\rho_j)\Bigr)+
\medskip \\ \displaystyle
+e^{-im\phi_j}~\Bigl(a_{j,-m}(\beta)
\mathbf{M}^2_{\omega\beta,-m}(\rho_j)-b_{j,-m}(\beta)
\mathbf{N}^2_{\omega\beta,-m}(\rho_j)\Bigr)\Bigr\}.
\end{array}
\label{G260}
\end{equation}

The expression for the magnetic field is analogous. Here the sum
over $\beta$ means the sum over the eigenvalues of longitudinal
wave vector of the array. We have added the argument $\beta$ to
partial amplitudes since they may be different for different
eigenmodes of the array.

In the first approximation,

\begin{equation}
\left(\begin{matrix} a_{jm}(\beta) \\ b_{jm}(\beta)
\end{matrix}\right)=
\left(\begin{matrix} \pm a_{j,-m}(\beta) \\ \mp b_{j,-m}(\beta)
\end{matrix}\right)=A_{jm}(\beta)~\mathbf{u}_{jm},
\label{G270}
\end{equation}

where the upper sign is for the first case and the lower sign for
the second case.

The vector $\mathbf{u}_{jm}$ doesn't depend on $\beta$. The
eigenvalues $\beta$ differ slightly, so one can suppose that
$\mathbf{M}^2_{\omega\beta,\pm
m}(\rho_j)\sim\mathbf{M}^2_{\omega\beta_{jm},\pm m}(\rho_j)$,
$\mathbf{N}^2_{\omega\beta,\pm
m}(\rho_j)\sim\mathbf{N}^2_{\omega\beta_{jm},\pm m}(\rho_j)$.
Introducing the notation

\begin{equation}
A_{jm}(z)=\sum\limits_\beta~e^{i\beta z}~A_{jm}(\beta),
\label{G275}
\end{equation}

one gets

\begin{equation}
\begin{array}{r} \displaystyle
\mathbf{E}(t,\mathbf{r})=e^{-i\omega t}~\sum\limits_{j=1}^N
~A_{jm}(z)~\Bigl\{\tilde{a}_{jm}~\Bigl(e^{im\phi_j}~\mathbf{M}^2_{\omega\beta_{jm}
m}(\rho_j)\pm
e^{-im\phi_j}~\mathbf{M}^2_{\omega\beta_{jm}-m}(\rho_j)\Bigr)+
\medskip \\ \displaystyle
+\tilde{b}_{jm}~\Bigl(e^{im\phi_j}~\mathbf{N}^2_{\omega\beta_{jm}
m}(\rho_j)\mp
e^{-im\phi_j}~\mathbf{N}^2_{\omega\beta_{jm}-m}(\rho_j)\Bigr)\Bigl\}.
\end{array}
\label{G280}
\end{equation}

From Eqs. (\ref{G250}) and (\ref{G275}) it follows

\begin{equation}
\left(i\frac{d}{dz}+\beta_{jm}\right)~A_{jm}(z)+
\gamma~\Bigl(A_{j-1,m}(z)+A_{j+1,m}(z)\Bigr)=0. \label{G290}
\end{equation}

This equation coincides to Eq. (\ref{Tradi}).


\section{Application for isofrequency curves calculation.}
\label{Sec:Application}

Consider an infinite array of identical waveguides. The optical
eigenmodes in this system possess the form of Bloch waves:

\begin{equation}
\begin{array}{c} \displaystyle
\mathbf{E}(t,\mathbf{r})=e^{-i\omega t+i\beta
z+ikx}~\mathbf{U}(\mathbf{r}),
\medskip \\ \displaystyle
\mathbf{H}(t,\mathbf{r})=e^{-i\omega t+i\beta
z+ikx}~\mathbf{V}(\mathbf{r}),
\end{array}
\label{J010}
\end{equation}

where $\mathbf{U}(\mathbf{r})$, $\mathbf{V}(\mathbf{r})$ are the
periodical functions relatively to the coordinate $x$. Here $k$ is
the transverse quasi wave vector belonging to the interval
$-\pi<k\leq\pi$ (here the period of the array is assumed to be
unit). For the fixed frequency $\omega$ the longitudinal wave
vector $\beta$ is connected with the transverse quasi wave vector
$k$, and the function $\beta(k)$ is the so-called isofrequency
curve.

For the field outside the waveguides Eq. (\ref{J010}) results in
the relations for the partial amplitudes

\begin{equation}
a_{jm}=a_m~e^{ikja}, \qquad b_{jm}=b_m~e^{ikja}. \label{J020}
\end{equation}

For the case of periodical array of identical waveguides the
scattering matrices for all the waveguides are the same. Besides,
the coupling coefficients $U_{jm}^{ln}(\omega,\beta)$ depend on
$j-l$. So the system (\ref{MainSyst}) takes the form

\begin{equation}
\hat{S}^{-1}_m(\omega,\beta)~ \left(\begin{matrix} a_{jm} \\
b_{jm}
\end{matrix}\right)+
\sum\limits_{l=-\infty}^{+\infty}~\sum\limits_n~
U_m^n\Bigl(\omega,\beta,~(l-j)a\Bigr)~ \left(\begin{matrix} a_{ln} \\
b_{ln} \end{matrix}\right)=0. \label{J030}
\end{equation}

Substituting (\ref{J020}) to (\ref{J030}), one obtains

\begin{equation}
\hat{S}^{-1}_m(\omega,\beta)~\left(\begin{matrix} a_m \\ b_m
\end{matrix}\right)+
\sum\limits_n~
U_m^n(\omega,\beta,k)~ \left(\begin{matrix} a_n \\
b_n \end{matrix}\right)=0, \label{J040}
\end{equation}

where

\begin{equation}
U_m^n(\omega,\beta,k)=\sum\limits_{l=-\infty}^{+\infty}~
U_m^n\Bigl(\omega,\beta,~(l-j)a\Bigr)~e^{ik(l-j)a}. \label{J050}
\end{equation}

In the matrix form this system of equations can be written as

\begin{equation}
\hat{L}(\omega,\beta,k)~\mathbf{x}=0, \label{J051}
\end{equation}

where the matrix $\hat{L}(\beta,k)$ contains the scattering
matrices $\hat{S}^{-1}_m(\omega,\beta)$ and the coupling
coefficients $U_m^n(\omega,\beta,k)$, and the column vector
$\mathbf{x}$ contains the partial amplitudes $a_m$, $b_m$.

The nontrivial solutions of system (\ref{J040}) exists when

\begin{equation}
\det\,\hat{L}(\omega,\beta,k)=0. \label{J055}
\end{equation}

For fixed frequency $\omega$ this equation implicitly determines
the isofrequency curves $\beta(k)$.

\bigskip

The isofrequency curves can be derived also from the
phenomenological model. Consider the equations (\ref{G070}) and
(\ref{G250}). Assuming $A_{jm}=e^{ikaj}~A_m$ and substituting this
to (\ref{G250}), one immediately gets an explicit expression for
isofrequency curves:

\begin{equation}
\beta(k)=\beta_m+2\gamma~\cos\,ka. \label{J056}
\end{equation}

Here $\beta_m$ is the propagation constant corresponding to the
angular momenta $m$ and $-m$ (it is independent on the number $j$
of a waveguide since all the waveguides are identical). The
similar result can be obtained from Eq. (\ref{G070}) after
substitution $a_{j0}=e^{ikaj}~a_0$.

As it was mentioned above, for $m\neq 0$ the system (\ref{G080})
possesses two types of solutions with two different coupling
constants $\gamma$. So, the propagation constant $\beta_m$, $m\neq
0$ gives rise to two different isofrequency curves. For the case
$m=0$ the propagation constant gives rise to one isofrequency
curve only.

\bigskip

Here we compare the isofrequency curves calculated by means of the
phenomenological model with the results of the rigorous model
based on the multiple scattering formalism. I. e. we compare the
results of calculations based on equations (\ref{J055}) and
(\ref{J056}).

We take the array of waveguides of refractive index
$n_{\text{wg}}=1.554$, and the refractive index of the medium
outside the waveguides is $n_{\text{med}}=1.457$. We suppose that
the period of the array is unit, $a=1$. The waveguides are
supposed to be situated close to each other, i. e. the radii of
the waveguides are $R=0.5$. The velocity of light in vacuum is
assumed $c=1$.

Below we use the multiple Mie scattering formalism for calculating
several isofrequency curves originating from different propagation
constants. The obtained isofrequency curves are compared with the
prediction of the phenomenological model. To calculate the
coupling constants $\gamma$ we use the formulae obtained in Sec.
\ref{Sec:Relationship}.

\bigskip

For angular momentum $m=0$ we take two propagation constants:
$\beta_1=126.671$ and $\beta_2=126.704$. The coupling constants
for them are $\gamma_1=-3.92\times 10^{-2}$ and
$\gamma_2=-3.78\times 10^{-2}$.

\begin{figure}[ptbh]
\centering
\includegraphics[width=0.7\textwidth] {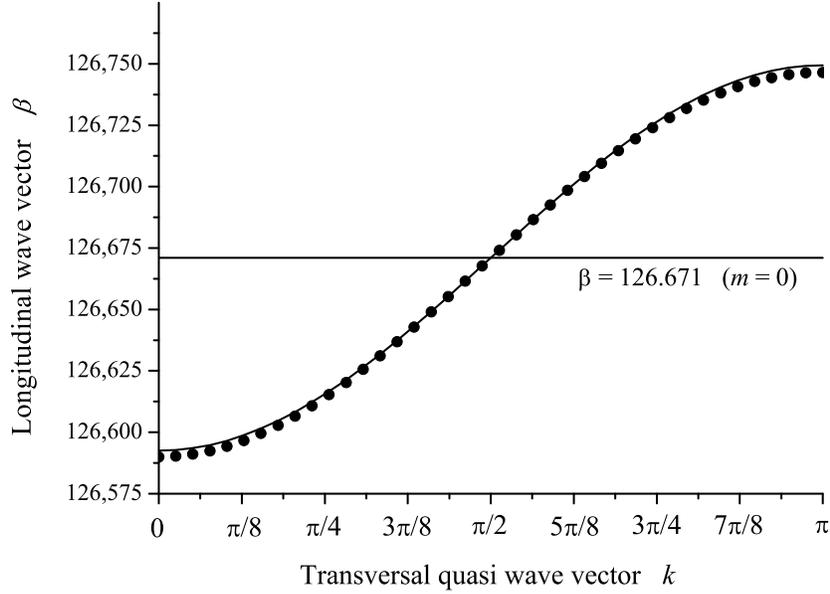}
\caption{Isofrequency curve originating from the propagation
constant $\beta=126.671$ ($m=0$). Dots for the curve obtained by
MMSF, solid line for the curve obtained by phenomenological
model.} \label{Fig1}
\end{figure}

\begin{figure}[ptbh]
\centering
\includegraphics[width=0.7\textwidth] {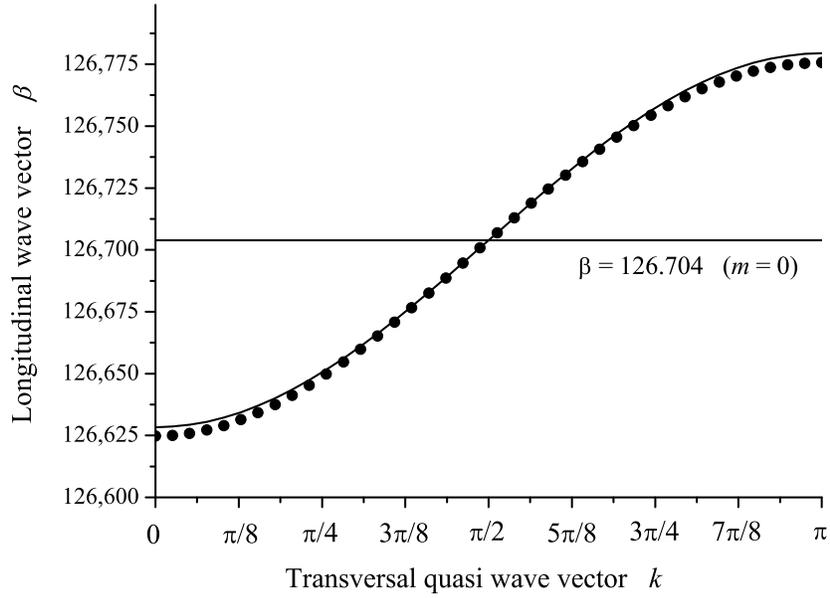}
\caption{Isofrequency curve originating from the propagation
constant $\beta=126.704$ ($m=0$). Dots for the curve obtained by
MMSF, solid line for the curve obtained by phenomenological
model.} \label{Fig2}
\end{figure}

The obtained isofrequency curves are presented in Fig. \ref{Fig1}
and Fig. \ref{Fig2}. The results of calculation by means of MMSF
are presented by dots, and the predictions of the phenomenological
model are shown by the solid curves. The horizontal lines show the
propagation constants. (Here and below the isofrequency curves are
plotted for $0<k<\pi$ since the function $\beta(k)$ is even,
$\beta(-k)=\beta(k)$.) One can see that the isofrequency curves
obtained by MMSF and the phenomenological model almost coincide.

\bigskip

For the angular momentum $m=1$ we take two propagation constants
also: $\beta_3=131.099$ and
$\beta_4=132.0092$. For every of propagation constants two
coupling constants exist. For $\beta_3$ the coupling constants are
$\gamma'_3=9.51\times 10^{-3}$ and $\gamma''_3=5.97\times
10^{-3}$. For $\beta_4$ they are
$\gamma'_4=7.90\times 10^{-4}$ and $\gamma''_4=7.12\times
10^{-4}$.

\begin{figure}[ptbh]
\centering
\includegraphics[width=0.7\textwidth] {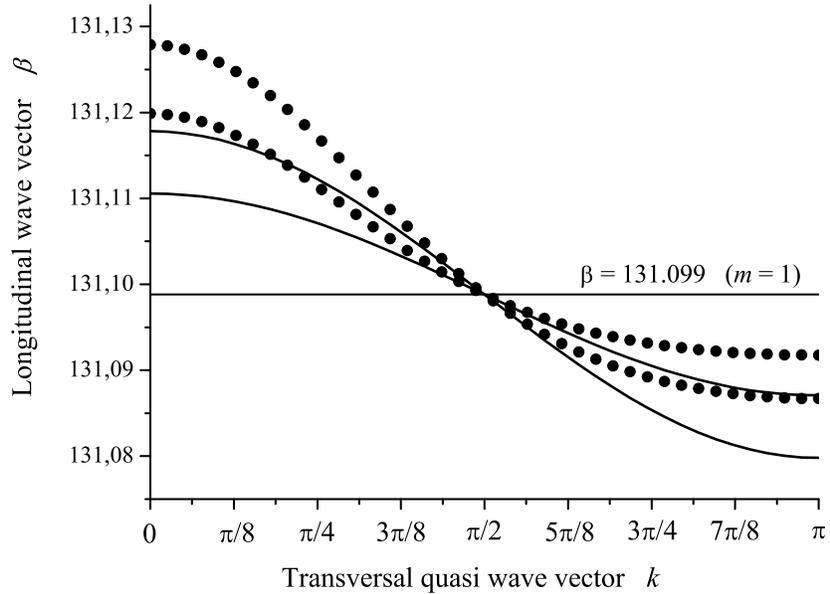}
\caption{Isofrequency curves originating from the propagation
constant $\beta=131.099$ ($m=1$). Dots for the curves obtained by
MMSF, solid lines for the curves obtained by phenomenological
model.} \label{Fig3}
\end{figure}

\begin{figure}[ptbh]
\centering
\includegraphics[width=0.7\textwidth] {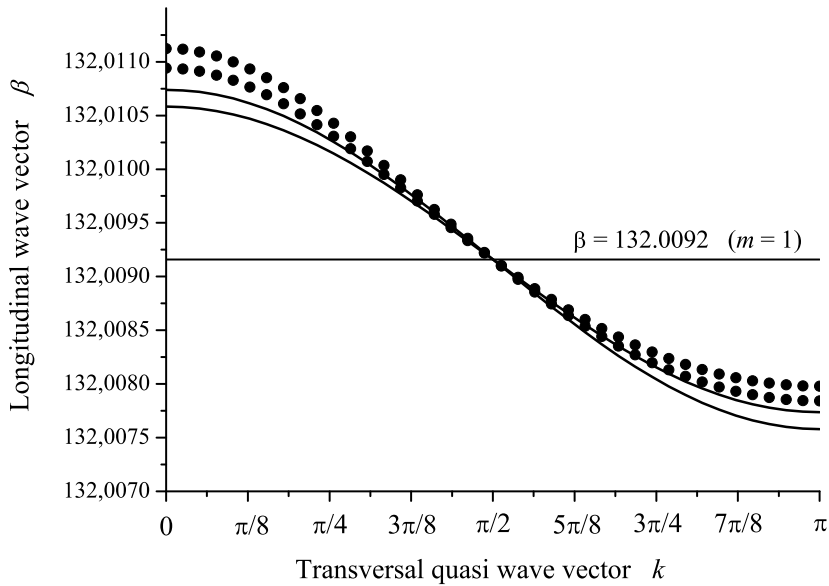}
\caption{Isofrequency curves originating from the propagation
constant $\beta=132.0092$ ($m=1$). Dots for the curves obtained by
MMSF, solid lines for the curves obtained by phenomenological
model.} \label{Fig4}
\end{figure}

The obtained isofrequency curves are represented in Fig.
\ref{Fig3} and Fig. \ref{Fig4}. One can see that the agreement
between the results of MMSF and phenomenological model for the
angular momentum $m=1$ is much worth then for $m=0$. In spite of
this, the phenomenological model is applicable for the qualitative
description of the isofrequency curves.

\newpage
\section{Conclusion.} \label{Sec:Conclusion}

In this paper we considered the planar arrays of cylindrical rods
by means of two methods: the phenomenological model and the
multiple Mie scattering formalism.

The MMSF has several advantages over the phenomenological method
based on Eq. (\ref{Tradi}). First, the MMSF allows to calculate
the behavior of the optical excitation for the case of the strong
coupling between the waveguides, while the phenomenological method
is applicable only for the case of the weak coupling. Second, the
input data for the MMSF are the geometrical properties of the
array and refractive indices of waveguides, while the
phenomenological method requires some data that should be obtained
experimentally, such as the propagation constants of waveguides
and coupling constants.

We demonstrated that for the case of evanescent coupling of rods
the phenomenological model can be derived from MMSF. We developed
the method to calculate the propagation constants $\beta_{jm}$ and
coupling constants $\gamma$. The applicability of the developed
method is demonstrated for different isofrequency curves. The
method represented in this work allows to produce the numerical
simulation without need of experimental investigation of
components of optical devices.

The method developed in this paper was used for isofrequency
curves calculation for the case of weak interaction between the
waveguides only. But it may be useful also for the systems with
strong coupling between the waveguides. In this case the
hybridization of modes with different angular momenta may take
place due to the coupling. Mathematically it means that one can't
neglect the coupling coefficients $U_{jm}^{ln}(\omega,\beta)$ with
$n\neq m$. In this situation the isofrequency curves may possess
the shape much more complicated than the phenomenological model
predicts.

The MMSF represented in this paper is convenient only for the
waveguides of cylindrical form, because in this case the
scattering matrix can be calculated easily. However, this method
can be applied for the waveguides of another shape, but in this
case it would be more difficult to calculate the scattering
matrix. Besides, the scattering by noncylindrical waveguides would
mix the harmonics with different angular momenta. Mathematically
it means that the scattering matrix $\hat{S}(\omega,\beta)$
contains some ``nondiagonal'' elements describing the transition
of harmonics $e^{im\phi_j}\,\mathbf{M}^1_{\omega\beta m}(\rho_j)$,
$e^{im\phi_j}\,\mathbf{N}^1_{\omega\beta m}(\rho_j)$ to harmonics
$e^{in\phi_j}\,\mathbf{M}^2_{\omega\beta n}(\rho_j)$,
$e^{in\phi_j}\,\mathbf{N}^2_{\omega\beta n}(\rho_j)$ with $n\neq
m$. Due to the existence of nonzero ``nondiagonal'' elements, the
scattering matrix $\hat{S}(\omega,\beta)$ can't be separated to
several matrices $\hat{S}_m(\omega,\beta)$ with fixed $m$.


\end{document}